\documentclass[aps,twocolumn,showpacs,preprintnumbers,amsmath,amssymb,nofootinbib]{revtex4}

\usepackage{amsmath,amssymb}
\def\s{\mbox{\boldmath$\displaystyle\mathbf{\sigma}$}}
\def\J{\mbox{\boldmath$\displaystyle\mathbf{J}$}}
\def\K{\mbox{\boldmath$\displaystyle\mathbf{K}$}}
\def\A{\mbox{\boldmath$\displaystyle\mathbf{A}$}}

\def\p{\mbox{\boldmath$\displaystyle\mathbf{p}$}}
\def\hp{\mbox{\boldmath$\displaystyle\mathbf{\widehat{\p}}$}}

\def\hp{\mbox{\boldmath$\displaystyle\mathbf{\widehat{\p}}$}}

\def\x{\mbox{\boldmath$\displaystyle\mathbf{x}$}}

\def\0{\mbox{\boldmath$\displaystyle\mathbf{0}$}}
\def\bv{\mbox{\boldmath$\displaystyle\mathbf{\varphi}$}}

\def\h00h{\mbox{\boldmath$\displaystyle\mathbf{(1/2,0)\oplus(0,1/2)}$}}

\def\rb{\kappa^+}
\def\lb{\kappa^-}

\newcommand{\ecco}{Elko}
\def\es{{\textit{\ecco~}}}

\newcommand{\mawave}{{\mathcal G}} 

\newcommand{\beqs}{\begin{equation*}}
\newcommand{\beq}{\begin{equation}}

\newcommand{\eeqs}{\end{equation*}}
\newcommand{\eeq}{\end{equation}}

\newcommand{\beqas}{\begin{eqnarray*}}
\newcommand{\beqa}{\begin{eqnarray}}

\newcommand{\eeqas}{\end{eqnarray*}}
\newcommand{\eeqa}{\end{eqnarray}}

\newcommand{\blist}{\begin{itemize}}

\newcommand{\elist}{\end{itemize}}

\newcommand{\Par}{$P$}
\newcommand{\Cha}{$C$}
\newcommand{\Tim}{$T$}

\begin{document}


\title{Dark matter: A spin one half fermion field with mass dimension one?}

\author{D.~V. Ahluwalia-Khalilova$^a$}
\email{dva-k@heritage.reduaz.mx} 

\author{D.~Grumiller$^b$} 
\email{grumiller@itp.uni-leipzig.de}

\affiliation{$^a$
ASGBG/CIU, Department of Mathematics, Ap. Postal C-600, 
University of Zacatecas,  ZAC 98060, Mexico \\
$^b$ Institut f\"ur Theoretische Physik, Universit\"at Leipzig, 
        Augustusplatz 10, D-04109 Leipzig, Germany}

\date{October 15, 2004}

\begin{abstract}
We report an unexpected theoretical discovery of a spin one 
half matter field with mass dimension one. It is based on a complete
set of eigenspinors of the charge conjugation operator.
Due to its unusual properties with respect to charge conjugation 
and parity it belongs to a non standard Wigner class. Consequently,
the theory exhibits non-locality with $(CPT)^2=-\mathbb{I}$.  Its dominant interaction with 
known forms of matter is via Higgs, and with gravity. 
This aspect leads us to contemplate it as a first-principle 
candidate for dark matter.
\end{abstract}


\maketitle

\section{Introduction}

The twentieth century may be described accurately as an era of 
local quantum field theories (QFTs). The concepts emerged in its 
first half, after unifying quantum mechanics, special relativity, 
and classical field theory. The applications were found and 
studied in detail, especially in the second half. This has culminated 
in the fantastically successful Standard Model (SM) of particle physics 
which describes all known forces of Nature except for gravity  
\cite{Weinberg:1995I}. 
As the unification of gravity with the quantum realm is still a 
work in progress it is worthwhile to tread gingerly. A safe, if somewhat vague, statement is that quantum gravity induces non-locality. 
This is realized in different ways explicitly in string theory \cite{Eliezer:1989cr}, 
in loop quantum gravity \cite{Rovelli:1998yv}, and in non-commutative field 
theories \cite{Douglas:2001ba}, 
to name just the most prominent candidates for quantum gravity. 
Clearly, abandoning locality is a big step. Therefore, we would like 
to be as conservative as possible regarding further deviations from 
the SM and its foundation in local QFT.

{\em Non standard Wigner classes.\textemdash}
Just dropping the postulate of locality is not specific
enough. The path we will
take is built upon the classic framework of Wigner \cite{Wigner:1939cj} where
particles are described by irreducible projective
representations of the {\em full} Poincar{\'e} group. At the kinematic 
level, they are
labeled by its
Casimir invariants.  In addition, they are endowed
with certain behavior under parity $P$ and charge
conjugation $C$ as distinguished by various Wigner classes. 
This notion of particles is a corner stone of any
description of the low energy regime that we are able to explore
experimentally (``low'' with respect to the Planck scale). Thus, we would like
to keep it, and advocate an \textit{ab initio} exploration of a {\em non
standard Wigner class} (NSWC). At this point two important facts are
recalled: (a) For the {\em standard Wigner classes}, the {\Par} and {\Cha}
anticommute for fermions and commute for bosons; this is true for all
particles of the SM; and, (b) Any non-trivial theory built upon a NSWC has
to be non-local \cite{Lee:1966tdw}. The second property is the reason
why the NSWCs are discarded normally. However, we regard it not as a
disadvantage but as a virtue, because non-locality is introduced in a
definite way with no free parameters apart from particle
properties.
For the sake of concreteness we shall
focus on spin one half.

{\em Neutrality.\textemdash} 
The concrete path we take is guided by the following observation.
Typically, what localizes the otherwise extended field configurations 
like solitons is a conserved (topological) charge \cite{Rajaraman:1982is}. 
In the absence of such charges there is nothing that 
protects the field from spreading; 
and therefore, in such situations, 
the emergence of non-locality may  not be completely surprising. 
We are thus led to consider Eigenspinors of the charge conjugation 
operator; abbreviated as \es from the German ``{\bf E}igenspinoren 
des {\bf L}adungs{\bf k}onjugations{\bf o}perators''. 
As we will show below, the assumption of neutrality alone will 
not only lead to a special type of non-locality with a certain mass dependence, but also to a NSWC with 
$[C,P]=0$. Thus, we need solely one postulate, namely 
that the new matter field we seek to describe is built upon
{\em \ecco}.

{\em The discovery.\textemdash} 
The constructed matter field, despite carrying spin one half,
is endowed with mass dimension one. This aspect, as we will argue, 
makes it a first-principle dark matter candidate.

{\em Preliminaries.\textemdash}
The derivation of the Dirac equation as presented, e.g.~in Ref. 
\onlinecite{Ryder:1996}, 
carries a quantum mechanical aspect
in allowing for the fact that the two Weyl spaces may carry a relative
phase; and concurrently a relativistic
element via the Lorentz transformation properties of the Weyl spinors.
In turn, the very existence of the latter depends on the existence of
two spacetime $SU(2)$s, with the following generators
of transformation:
$ 
\A_\pm= \frac{1}{2}\left(\J\mp i\K\right)
$.
The $\J$ and $\K$ represent the generators of rotations and boosts, respectively. 
We use the Pauli matrices $\s=(\sigma_1,\sigma_2,\sigma_3)$ and the Dirac matrices $\gamma^\mu$ in standard Weyl representation, subsequently.
For $\J=\s/2$ and $\A_+=\0$ [$\A_-=\0$] we have the 
$(1/2,0)$ right-handed [$(0,1/2)$ left-handed]
Weyl space where $\K$ equals $-i \s/2$ [$ + i \s/2$]. 
From the womb of this structure emerges
the Dirac equation, $\left(\gamma^\mu p_\mu\pm m \mathbb{I}\right)\psi(\p)=0$,
which carries the particle-antiparticle symmetry via the
operation 
of charge conjugation. In Weyl realization, the operator 
associated with it is 
\beq
{C} = 
\left(
\begin{array}{cc}
\mathbb{O} & i\,\Theta \\
-i\,\Theta & \mathbb{O}
\end{array}
\right) {K}\,,\label{cc}
\eeq
where $K$ complex conjugates a spinor appearing 
on its right 
and  $\Theta$ is Wigner's spin half time reversal 
operator. We employ the representation $\Theta=-i\sigma_2$.
Note that $\Theta \left[\s/2\right] \Theta^{-1} = -\, \left[\s/2\right]^\ast$.
Equation~(\ref{cc}) yields 
the expected $C=-\gamma^2 K$.
The boost
operator, $\rb\oplus \lb$, with
\begin{equation}
\kappa^\pm =\exp\left(\pm\,\frac{\s}{2}\cdot\bv\right)= \sqrt{\frac{E+m}{2\,m}}
\left(\mathbb{I}\pm\frac{\s\cdot\p}{E+m}\right)
\,,  \label{br}
\end{equation}
and the 
$(1/2,0)\oplus(0,1/2)$-space charge conjugation 
operator, $C$, commute. In terms of energy $E$ and momentum $\p=p\,\hp$ the boost parameter, $\bv=\varphi\,\hp$,  is defined as
$\cosh(\varphi)=E/m$, $\sinh(\varphi)=p/m$, where $m$ is the mass. 

 
\section{Formal structure of \ecco}
We have summarized above the origin and form of {\Cha}. We now proceed
to obtain its eigenspinors. 
If $\phi_L(\p)$ 
transforms as a 
left 
handed spinor, then
$\left(\zeta \Theta\right) \,\phi_L^\ast(\p)$ 
transforms as a right 
handed spinor --- where 
$\zeta$ is
an unspecified phase.
As a consequence, the following spinors 
belong to the $(\frac{1}{2},0)\oplus(0,\frac{1}{2})$
representation space \footnote{There is a second set of spinors that may be built by starting with a right-handed Weyl spinor $\phi_R(\p)$, and the observation  that $(\zeta \Theta)^\ast \phi^\ast_R(\p)$ transforms as a left-handed Weyl spinor. Due to its {\em equivalence} with the set considered in the present work we postpone its details to \cite{Ahluwalia-Khalilova:2004ab}.}:
\begin{equation}
\lambda(\p) =
\left(
\begin{array}{c}
\left(\zeta \Theta\right) \,\phi_L^\ast(\p)\\
\phi_L(\p)
\end{array}
\right)
\end{equation}
These become Eigenspinors of {\Cha}, viz., {\em \ecco}, 
with real eigenvalues
if  the phase $\zeta$ 
is restricted to 
$\zeta= \pm\,i$: 
\beq
{C} \lambda(\p) = \pm  \lambda(\p) \label{eq:neutral}
\eeq
The plus [minus] sign 
yields  \textit{self-conjugate [anti self-conjugate] spinors: 
$\lambda^S(\p)$ [$\lambda^A(\p)$].}

{\em Explicit form of \ecco.\textemdash}
To obtain explicit expressions for $\lambda (\p )$ we consider the
 rest frame ($\p=\0$) and decompose the $\phi_L(\0)$ into helicity eigenstates:
$
\s\cdot{\hp} \;\phi_L^\pm (\0)= \pm\;\phi_L^\pm(\0)
$.
Taking $\hp = \left(\sin{\theta}\cos{\phi},\,
\sin{\theta}\sin{\phi},\,\cos{\theta}\right)$,
yields
\begin{subequations}
\beqa
&& \phi_L^+(\0) =
\sqrt{m} e^{i\vartheta_1} 
\left(
\begin{array}{c}
\cos(\theta/2) e^{-i\phi/2}\\
\sin(\theta/2) e^{i\phi/2}
\end{array}
\right)\,,\\
&& \phi_L^-(\0) =
\sqrt{m} e^{i\vartheta_2} 
\left(
\begin{array}{c}
\sin(\theta/2) e^{-i\phi/2}\\
-\cos(\theta/2) e^{i\phi/2}
\end{array}
\right)\,.
\eeqa
\end{subequations}
We set $\vartheta_1= \vartheta_2 =0$ \footnote{This choice 
is important for 
the specific norms given in Eqs. (\ref{zd1}) and (\ref{z5}).}. 
This leads to \textit{four} spinors 
\beq
\lambda_{\{\mp,\pm\}}(\0) = 
\left(
\begin{array}{c}
\zeta \,\Theta \,\left[\phi^\pm_L(\0)\right]^\ast\\
\phi^\pm_L(\0)
\end{array}
\right)\,.\label{restspinors}
\eeq
Two of these are [anti] self conjugate and arise from setting $\zeta= +i$ [$\zeta= -i$]. These are denoted by $\lambda^S_{\{\mp,\pm\}}(\0)$ [$\lambda^A_{\{\mp,\pm\}}(\0)$].
The first [second] helicity entry refers to the $(\frac{1}{2},0)$ 
[$(0,\frac{1}{2})$] transforming component of the
 $\lambda(\p)$.  
Equations (\ref{br}) and 
(\ref{restspinors}) yield the boosted spinors: 
\beq
\lambda_{\{\mp,\pm\}}^{S/A}(\p)
=
\sqrt{\frac{E+m}{2\,m}}\left(1\mp\frac{ p}{E+m}\right)
\lambda_{\{\mp,\pm\}}^{S/A}(\0) \label{lsup}
\eeq
In the massless limit $\lambda_{\{-,+\}}^{S/A}(\p)$ {\em identically vanishes} while $\lambda_{\{+,-\}}^{S/A}(\p)$ does not. Moreover, the relation,
$
\s\cdot\hp \, \Theta \left[\phi_L^\pm (\0)\right]^\ast
= \mp\, \Theta\left[\phi_L^\pm(\0)\right]^\ast
$,
physically implies the following: $\Theta \left[\phi_L^\pm (\0)\right]^\ast$
has opposite helicity of $\phi_L^\pm (\0)$.
Since $\s\cdot\hp$ commutes with 
$\kappa^\pm$ this result holds for all $\p$. 
We thus have the important property for {\em \ecco}: 
they
are {\em not\/} single helicity objects. 
That is, \es
cannot be eigenspinors of the helicity operator. 
The same shall be assumed for one-particle states.

{\em A new dual for \ecco.\textemdash}
For any $(\frac{1}{2},0)\oplus(0,\frac{1}{2})$ spinor  $\xi(\p)$, 
 the 
Dirac dual spinor $\overline{\xi}(\p)$ is defined as $\overline{\xi}(\p) := \xi^\dagger(\p) \gamma^0$.
It is readily verified that with respect to the Dirac dual, the 
\es have an imaginary  bi-orthogonal norm, which
is a hindrance to physical interpretation and quantization.
Therefore, we define a new dual 
which is required to have the 
property that: (a) It yields an invariant real definite norm, and
(b) It 
must secure a  positive definite norm for two of
the four  \es\hspace{-5pt}'s, 
and negative
definite norm
for the remaining two. 
 Up to an irrelevant relative 
sign, a unique 
definition, which we call 
{\em \es dual,\/} is
\beq
 \stackrel{\neg}\lambda^{S/A}_{\{\mp,\pm\}}(\p)
:= \pm \,i\,\left[\lambda^{S/A}_{\{\pm,\mp\}}(\p)\right]^\dagger\gamma^0\,.
\eeq
With the \es dual thus defined, we now have, by construction, 
the orthonormality relations 
\begin{subequations}
\begin{align}
& \stackrel{\neg}\lambda^S_{\alpha}(\p)\;
\lambda^I_{\alpha^\prime}(\p) = +\; 2 m\; \delta_{\alpha\alpha^\prime}\;\delta_{SI}\,,
\label{zd1}\\
& \stackrel{\neg}\lambda^A_{\alpha}(\p)\;
\lambda^I_{\alpha^\prime}(\p) = -\; 2 m \;\delta_{\alpha\alpha^\prime}\;\delta_{AI}\,,
\label{z5}
\end{align}
\end{subequations}
where $I\in\{S,A\}$;
and the completeness relation
\beq
\frac{1}{2 m}\sum_\alpha 
 \left[\lambda^S_{\alpha}(\p) \stackrel{\neg}\lambda^S_{\alpha}(\p) 
      - \lambda^A_{\alpha} (\p)\stackrel{\neg}\lambda^A_{\alpha}(\p)\right]
 = \mathbb{I}\,, \label{z1}
\eeq
which clearly shows the necessity of the anti self-conjugate
spinors.
In the above equations, the subscript $\alpha$ ranges  over two
possibilities: $ \{+,-\}, \{-,+\}$. 
 The detailed structure underlying the 
completeness relation resides in the following \textit{spin sums}
\begin{subequations}
\beqa
&&  \sum_{\alpha} \lambda^S_\alpha(\p)\stackrel{\neg}\lambda_\alpha^S(\p) 
= +\, m\big[ \mathbb{I}+\mawave(\p)\big]\,, \label{spinsumS}\\
&&  \sum_{\alpha} \lambda^A_\alpha(\p)\stackrel{\neg}\lambda_\alpha^A(\p) 
=  - m\,\big[\mathbb{I}-\mawave(\p)\big]\,; \label{spinsumA}
\eeqa
\end{subequations}
which together \textit{define} $\mawave(\p)$. 
A detailed calculation shows that $\mawave$
is an odd function of $\p$:
\beq
\mawave(\p)= - \,\mawave(- \p)\,;\label{eq:g}
\eeq 
a result which carries 
considerable significance 
for the discussion following Eq. (\ref{eq:noteworthy}).
Equations (\ref{zd1})-(\ref{spinsumA}) have their direct counterparts
in Dirac's construct.


{\em Behavior under {\Cha}, {\Par} and {\Tim}.\textemdash}
It appears to be standard textbook wisdom
that for bosons [fermions] particle and antiparticle have same [opposite] relative intrinsic parity.
To our knowledge the only textbook which tells a more intricate story is
that by Weinberg \cite{Weinberg:1995I}. The only known
explicit construct of a theory which challenges the 
conventional wisdom was reported about a decade ago 
\cite{Ahluwalia:1993zt}. In that pure spin one bosonic theory particles 
and antiparticles carry opposite, rather than same, relative 
intrinsic parity. 
It manifests itself 
through anticommutativity, as opposed to commutativity, 
of the $(1,0)\oplus(0,1)$-space's 
{\Cha} and {\Par} operators. 
In a somewhat parallel fashion we shall now show that 
for the spin half \es  {\Cha}
and {\Par} commute, rather than anticommute as they do
for the Dirac case. The {\Par} acting on \es
yields
\begin{equation}
P\lambda^{S}_{\{\mp,\pm\}} (\p)= \pm\, i\, \lambda^A_{\{\pm,\mp\}}(\p)\,,\label{eq:parity} 
\end{equation}
and the same equation with a minus sign on the l.h.s.~for $\lambda^A\leftrightarrow\lambda^S$.
That is, \es are \textit{not} eigenspinors of the 
parity operator. Applying it twice establishes
$
P^2=-\,\mathbb{I}
$;
as opposed to Dirac spinors where $P^2=+\mathbb{I}$. Under time reversal {\Tim$=i\gamma^5$\Cha} we obtain
\begin{equation}
\label{eq:time}
T \lambda_\alpha^S(\p) = - i \lambda_\alpha^A(\p),\quad  
T \lambda_\alpha^A(\p) = + i \lambda_\alpha^S(\p)\,,
\end{equation}
implying $T^2=-\mathbb{I}$. It is now a simple exercise to show 
\beq
\mbox{\sc \ecco}: \quad
[C,P]=0\,,\quad [C,T]=0\,,\quad \{P,T\}=0\,.
\nonumber
\eeq 
This proves our claim that \es belong to a NSWC \cite{Wigner:1939cj}. 
We confirm also Wigner's expectation $(CPT)^2=-\mathbb{I}$ and reconcile
with Weinberg's observation (appendix C of chapter 2 in \cite{Weinberg:1995I}) due to {\em \ecco}'s dual helicity nature.

\section{Physical properties of \ecco}

{\em A spin one half matter field with mass dimension one.\textemdash}
An {\em \ecco}-based quantum
field with well-defined $CPT$ properties may now be introduced 
\begin{multline}
 \eta(x)=\int \frac{d^3 p}{(2\pi)^3}\frac{1}{\sqrt{2\, m\, E(\p)}} 
\sum_{\beta}   \Big[c_\beta(\p) \lambda^S_\beta(\p) 
\mathrm{e}^{-ip_\mu x^\mu} \\
+ c_\beta^\dagger(\p)  
\lambda^A_\beta(\p) \mathrm{e}^{+ip_\mu x^\mu}\Big]\,,\label{eq:4.36ap} 
\end{multline}
with the expected anti-commutation relations
\begin{align}
& \big\{c_\beta(\p),\; c^\dagger_{\beta^\prime}(\p^\prime)\big\}
=  \left(2\pi\right)^3  \delta^3\left(\p-\p^\prime\right)
\delta_{\beta\beta^\prime}\,,\label{eq:anticomm} \\
&\big\{c^\dagger_\beta(\p),\; c^\dagger_{\beta^\prime}(\p^\prime)\big\}
=\big\{c_\beta(\p),\; c_{\beta^\prime}(\p^\prime)\big\}=0\,,
\label{eq:anticommz}
\end{align}
for the creation and annihilation operators $c^\dagger_\beta(\p)$
and $c_\beta(\p)$, respectively.
Its \es dual  ${\stackrel{\neg}\eta}(x)$  
is obtained by replacing everywhere $\lambda(\p)$ 
with its \es dual, exchanging $c$ with $c^\dagger$, and swapping $ip_\mu x^\mu\leftrightarrow-ip_\mu x^\mu$. 
The propagator follows from textbook methods. 
It entails evaluation of 
$\langle \hspace{8pt}\vert T (\eta(x^\prime)\stackrel{\neg}
\eta(x))\vert\hspace{8pt}\rangle$, where $T$ is
the fermionic time-ordering operator, and
$\vert\hspace{8pt}\rangle$ is the vacuum state. 
The result in terms of the spin sums reads 
\beqa
 &&\hspace{-36pt}\mathcal{S} (x-x^\prime) = -
 \int \frac{d^3p}{(2\pi)^3}\;
\frac{i}{2 m E(\p)} \nonumber\\ 
&& \times\sum_\beta \Big[ \theta(t^\prime-t)
 \lambda_\beta^S(\p) \stackrel{\neg}\lambda^S_\beta(\p)
\mathrm{e}^{- i p_\mu(x^{\prime\mu} - x^\mu)} \nonumber\\
&&    -\;
 \theta(t-t^\prime) \lambda_\beta^A(\p) \stackrel{\neg}\lambda^A_\beta(\p)
\mathrm{e}^{+ i p_\mu(x^{\prime\mu} - x^\mu)}
\Big]\,.\label{eq:6.a}
\eeqa
On using  Eqs. (\ref{spinsumS}) and (\ref{spinsumA}) for the spins sums 
it simplifies to 
\beq
\mathcal{S} (x-x^\prime) = \int\frac{d^4 p}{(2\pi)^4}\;
\mathrm{e}^{i p_\mu(x^\mu-x^{\prime\mu})}\;  
\frac{\mathbb{I}+\mawave(\p)}{p_\mu p^\mu - m^2 + i\epsilon}\,.
\label{eq:noteworthy}
\eeq
In \eqref{eq:noteworthy}, the limit $\epsilon\to0^+$ is understood.
The structure of the obtained propagator differs from that of Dirac
because in this latter case $\left(\mathbb{I}\pm\mathcal{G}(\p)\right)$ 
appearing in the spin
sums is replaced by its counterpart 
$- \left( \mathbb{I} \pm \gamma_\mu p^\mu/m\right)$ (with plus sign giving
the spin sum for particle spinors $u_h(\p)$, 
while the minus sign yielding the same for
antiparticle spinors $v_h(\p)$). 
Exploiting the property (\ref{eq:g}),  
it is clear that in the absence of a preferred direction, such as the one arising from a
fixed background, like a reference fluid, a thermal bath or an 
external magnetic field, to name just a few, 
the second term in Eq.~(\ref{eq:noteworthy}) identically vanishes;  
and as a result, Eq.~\eqref{eq:noteworthy} reduces to
the Klein-Gor\-don pro\-pa\-ga\-tor.  Consequently, the 
field $\eta(x)$ carries mass dimension one. It forbids particles
described by the theory to enter $SU(2)_L$ doublets of the 
SM. The field $\eta(x)$ thus becomes
a first-principle candidate for dark matter as will be discussed below in more detail.

The identity \footnote{Here
$\delta^\alpha_\beta$ is the usual Kronecker delta, 
the antisymmetric symbol is defined as 
$\varepsilon^{\{-,+\}}_{\{+,-\}}:=-1$, 
and $+(-)$ sign is to be taken for self-conjugate (anti self-conjugate)
spinors.} 
$\left(\gamma_\mu p^\mu \delta_\alpha^\beta \pm im\mathbb{I}
\varepsilon_\alpha^\beta\right)\lambda_\beta^{S/A}(\p)=0,$
follows as a simple algebraic exercise of
applying $\gamma_\mu p^\mu$ to $\lambda^{S/A}(\p)$ 
 \cite{Dvoeglazov:1995kn,Ahluwalia-Khalilova:2004ab}.
It \textit{cannot} be interpreted
as Dirac equation with an off-diagonal mass term.
Instead, the mentioned identity
shows that {\em \ecco} satisfy the 
Klein-Gordon equation, $(p_\mu p^\mu-m^2)\lambda^{S/A}(\p)=0$. 

As a further  consistency check, from the Lagrangian density
\begin{equation}
  \label{eq:lagrangiandensity}
  \mathcal{L}^{\rm free} = \partial^\mu {\stackrel{\neg}\eta}(x) \,\partial_\mu\eta(x) - m^2\stackrel{\neg}\eta(x)\,\eta(x)\,,
\end{equation}
one may construct the Hamiltonian density and it turns out that the anti-commutation relations \eqref{eq:anticomm}, \eqref{eq:anticommz} are compatible with positive energy, like in the Dirac case.

{\em Non-locality.\textemdash} 
Given \eqref{eq:4.36ap}, its dual $\stackrel{\neg}{\eta}(x)$, 
as well as 
the canonical momentum $\pi(x)$ implied by Eq.~(\ref{eq:lagrangiandensity}),
as input it is easy to calculate the field anti-commutators. 
We find that 
$\big\{\eta(\x,t),\,\stackrel{\neg}{\eta}(\x^\prime,t)\big\} $ vanishes while
$\big\{\eta(\x,t),\,\pi(\x^\prime,t)\big\}= i\delta^3(\x-\x^\prime)$. This is
as expected on the basis of a local QFT.
The departure from locality is contained in the result  that
$\{\eta(\x,t),\eta(\x^\prime,t)\} $ and 
$\{\pi(\x,\,t),\;\pi(\x^\prime,\,t)\}$ do \textit{not} vanish.
The emergent non-locality is captured by the expression 
\begin{equation}
  \label{eq:nonlocren}
  \frac{\mathrm{d}}{\mathrm{d}m} 
\left[m \int_{\x-\x^\prime}\Big\langle\hspace{11pt}
\Big\vert\left\{\eta(\x,t),\eta(\x^\prime,t)\right\}
\Big\vert\hspace{11pt}\Big\rangle\right] = \frac{1}{m}
\gamma^1\gamma^0 \,. \nonumber  
\end{equation}
In the limit of large $m$ non-locality becomes negligible.
It is worth emphasizing that
non-locality for \es emerges as a higher order effect; 
for it resides entirely in those expectation values
where  two \es fields, or two momenta, appear together.

{\em {\ecco} as dark matter candidate.\textemdash}
Having established non-locality, $CPT$-properties and mass dimension one, 
the physics of \es becomes even more interesting when coupling to the matter content of the SM is considered.
Since interaction terms with mass dimension greater than four will be assumed to be suppressed by some fundamental mass scale, say, the Planck scale, focus will be solely on power counting renormalizable and super-renormalizable terms \footnote{We do not intend to discuss renormalizability which is tricky for non-local theories, but rather impose only simple power counting arguments in order to extract the dominant terms in the low energy limit.}. It is easy to check that none of the latter are present: for a scalar interaction term \es must appear in even powers, so super-renormalizable terms must contain exactly two {\ecco}s and one other field. However, it cannot be a spinor (or else the interaction term would not be a scalar) or a gauge-field (or else the interaction term would not be gauge invariant). Therefore, only a {\em neutral} scalar field remains as possible candidate. The only scalar field within the SM is the Higgs, which is an $SU(2)_L$ doublet. Thus, only power counting renormalizable terms have to be considered. In addition to the free Lagrangian density \eqref{eq:lagrangiandensity} and quartic \es self-interactions there is a possible {\em \ecco}-Higgs interaction 
\begin{equation}
 \label{eq:higgs}  
\mathcal{L}^{\rm H} = \alpha_H \,\phi^2(x) \stackrel{\neg}{\eta}(x)\eta(x)
\,, 
\end{equation}  
where $\phi(x)$ is the Higgs doublet and $\alpha_H$ is a dimensionless 
coupling constant.
The fact that \es may not interact directly with non-abelian gauge fields 
\footnote{While \es may carry a coupling to an abelian gauge field with associated field strength $F_{\mu\nu}(x)$, e.g.~of the form 
$\stackrel{\neg}{\eta}(x)\gamma^\mu\gamma^\nu F_{\mu\nu}(x)\eta(x)$, 
the coupling constant has to be very small because such terms affect photon propagation. Thus, the dominant interaction between \es and particles of the SM is expected to be via \eqref{eq:higgs}. We thank Dima Vassilevich for raising a question in this regard.} 
or fermions of the SM explains why \es has not been detected yet. 
However, since it does interact with the Higgs there is a chance that it 
might be discovered at LHC.
Thus, due
to its weak interaction with the matter content of the SM \es provides
 a first-principle candidate for dark matter.


{\em Conclusions.\textemdash}
Perhaps it not too provocative an assertion that 
whatever dark matter is, one thing that seems reasonably 
assured is that in the low-energy
limit it behaves as one  of  the representations of the Lorentz group.
Since the known particles are described by quantum fields involving finite
 dimensional representation spaces, and since none of them fits the
properties
 called for by dark matter, one is guided to study the matter content of
 the unexplored Wigner classes. Here, we have examined one such spin one
 half representation space. It is emphasized that all our findings depend crucially on a single postulate: neutrality, as encoded in Eq.~\eqref{eq:neutral}.

Not  only do our results offer a possible new
 candidate for dark matter, but they also provide unexpected theoretical
 insights into the particle content of the spacetime symmetries.

 
\acknowledgments

We are grateful to Terry Pilling 
and Dima Vassilevich for helpful discussions.
CONACyT (Mexico) is  acknowledged 
for funding this research through Project
32067-E.
D. Grumiller is supported by 
project J-2330-N08 of the Austrian 
Science Foundation (FWF).

\bigskip

{\em Note added.\textemdash} 
 During the time this paper was under review Ref.~\cite{roro} appeared.
 In that paper da Rocha and Rodrigues calculate the bilinear
 covariants for the Elko spinor fields and show that Elko belongs
 to class 5 in Lounesto spinor classification \cite{loun}. They further
 discuss distinction between Elko and Majorana spinors. In addition,
 if  Elko is to serve as a dark matter candidate in the standard
 model of cosmology,  Ref.~\cite{Ahluwalia-Khalilova:2004ab} provides an estimate for the Elko mass
 (about 20 MeV) and the relevant cross section (roughly 2pb).
 A refinement of that analysis in the form of an S-matrix calculation is
 desirable. First steps in this direction are also provided in Ref. \cite{Ahluwalia-Khalilova:2004ab},
 where the impact of non-locality on a perturbative treatment has been
 studied to a certain extent. In particular, non-standard contractions
 emerge in the analogue of Wick's theorem.




\end{document}